# A Guide to Design Disturbance Observer-based Motion Control Systems in Discrete-time Domain

Emre Sariyildiz
School of Mechanical, Materials, Mechatronic and Biomedical Engineering,
University of Wollongong, NSW, 2522, Australia.
emre@uow.edu.au

*Abstract*- **This paper analyses and synthesises the Disturbance Observer (DOb) based motion control systems in the discrete-time domain. By employing Bode Integral Theorem, it is shown that continuous-time analysis methods fall-short in explaining the dynamic behaviours of the DOb-based robust motion controllers implemented by computers and microcontrollers. For example, continuous-time analysis methods cannot explain why the robust stability and performance of the digital motion controller deteriorate as the bandwidth of the DOb increases. Therefore, unexpected dynamic responses (e.g., poor stability and performance, and high-sensitivity to disturbances and noise) may be observed when the parameters of the digital robust motion controller are tuned by using continuous-time synthesis methods in practice. This paper also analytically derives the robust stability and performance constraints of the DOb-based motion controllers in the discrete-time domain. The proposed design constraints allow one to systematically synthesise a high-performance digital robust motion controller. The validity of the proposed analysis and synthesis methods are verified by simulations.**

***Index Terms: Discrete-time Control, Disturbance Observer, Motion Control, Robust Control, Robust Stability and Performance.***

## I. INTRODUCTION

It is a well-known fact that the stability and performance of a motion control system may significantly deteriorate by internal and external disturbances (e.g., friction, parametric uncertainties and unknown dynamics in plant model, and load) in real world applications [1–5]. To deal with this problem, various adaptive and robust motion control techniques, such as Sliding Mode Control, H∞ control, Internal Model Control and Robust Parametric Control, have been proposed in the last decades [4–9]. Among them, the DOb is one of the most widely used robust motion control tools due to its simplicity and efficacy [1, 10].

The DOb-based robust motion controller synthesis is based on a simple and elegant idea: cancelling disturbances via feedforward control. The internal and external disturbances of a motion control system are cancelled by feedforwarding the reverse of the disturbance signal [1]. This intuitive robust motion controller design technique has attracted many control engineering practitioners, and the DOb has been applied to various engineering systems (e.g., robots, hard-disk drives, automobiles, satellites and unmanned aerial vehicles) in the last three decades [1, 11–15].

However, the feedforward robust controller synthesis is impractical because the internal and external disturbances are generally unknown in motion control systems [1]. In practice, the disturbances are estimated by using the measureable states and known plant dynamics, i.e., DOb, and the robust motion controller is synthesised by feedforwarding the reverse of the estimated disturbance signal. In other words, a feedback robust motion controller is implicitly synthesised [1, 16, 17]. The stability and performance of the robust motion controller are directly influenced by the dynamics of disturbance estimation, i.e., the design parameters of the DOb [18]. For example, it is a well-known fact that the stability of the DOb-based robust motion controller deteriorates as the nominal inertia decreases [16, 17]. It is therefore very important to understand how the dynamic response of the robust motion controller changes by the design parameters of the DOb.

To improve the robust stability and performance of the DOb-based motion control systems, several analysis and synthesis methods have been proposed in the literature [16–20]. Although computers and microcontrollers are always employed in the implementation of the robust motion controllers, continuous-time analysis methods are generally used due to simplicity [1, 21]. However, continuous-time analysis methods cannot explain all dynamic responses, such as poor stability and performance, of the DOb-based robust motion controller implemented by computers and/or microcontrollers. For example, it is shown that the DOb-based digital robust motion control systems may exhibit under-damped and even unstable responses as the bandwidth of disturbance estimation increases in [22–24]. In section III, this paper clarifies why continuous-time analysis methods fall-short in explaining the dynamic behaviours of the DOb-based digital robust motion controllers by employing Bode Integral Theorem [25]. The following studies analyse and synthesise the DOb-based robust motion controllers in the discrete-time domain: bilinear transformation is used in [26, 27], sensitivity optimisation method is used to obtain better tracking performance than bilinear transformation in [28], disturbance suppression is improved by using multi-rate sampling control method in [29], optimal plant models are proposed to improve the bandwidth of disturbance estimation in [30, 31], Kalman filter is combined with the DOb to improve disturbance estimation in [32]. Nevertheless, the robust stability and performance of the DOb-based digital robust motion controllers have not been discussed in detail [24, 33]. Moreover, the design constraints of the DOb-based motion control systems have not yet been derived in the discrete-time domain.

This paper proposes a guide to design the DOb-based digital robust motion control systems. The design constraints of the digital robust motion controller (i.e., the bandwidth of the DOb, nominal plant model and sampling frequency) are analytically derived in discrete-time. The proposed design constraints allow one to systematically synthesise a high-performance robust digital motion controller. Bode Integral Theorem is employed in the continuous- and discrete- time

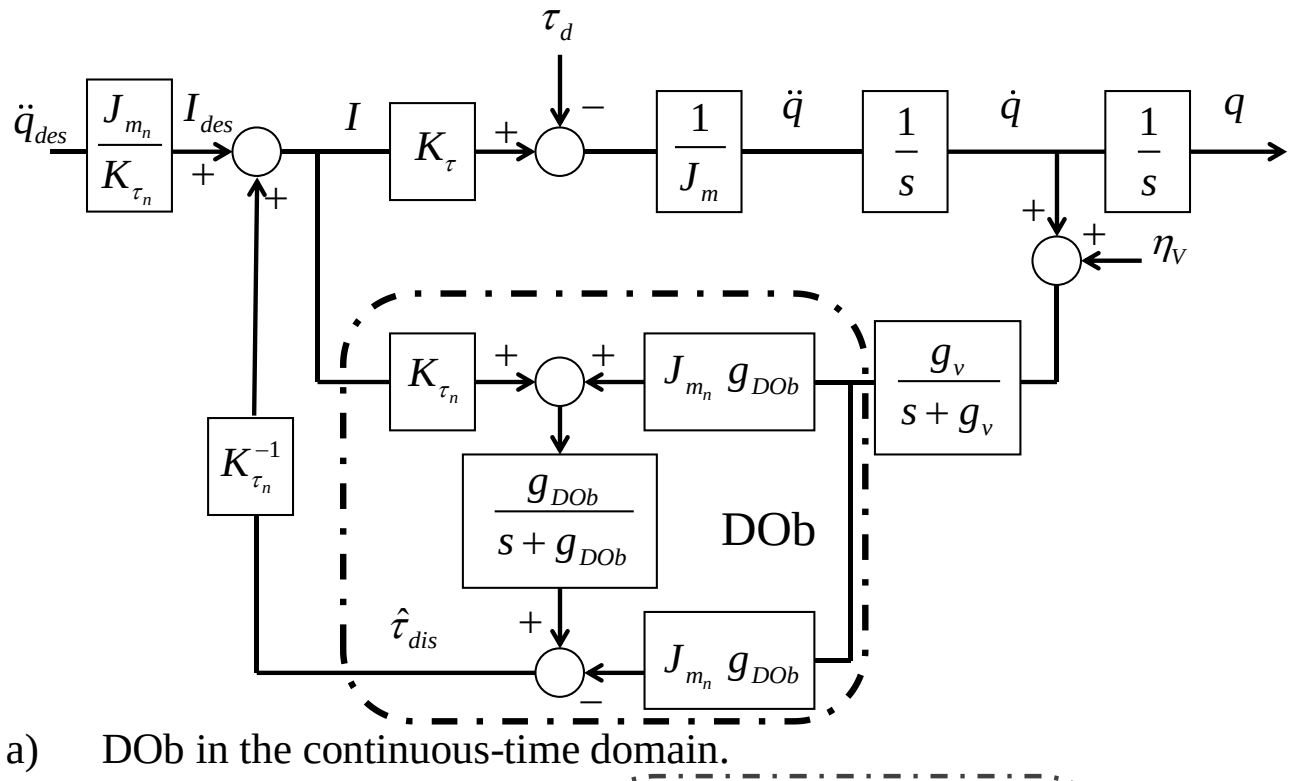

a) DOb in the continuous-time domain.

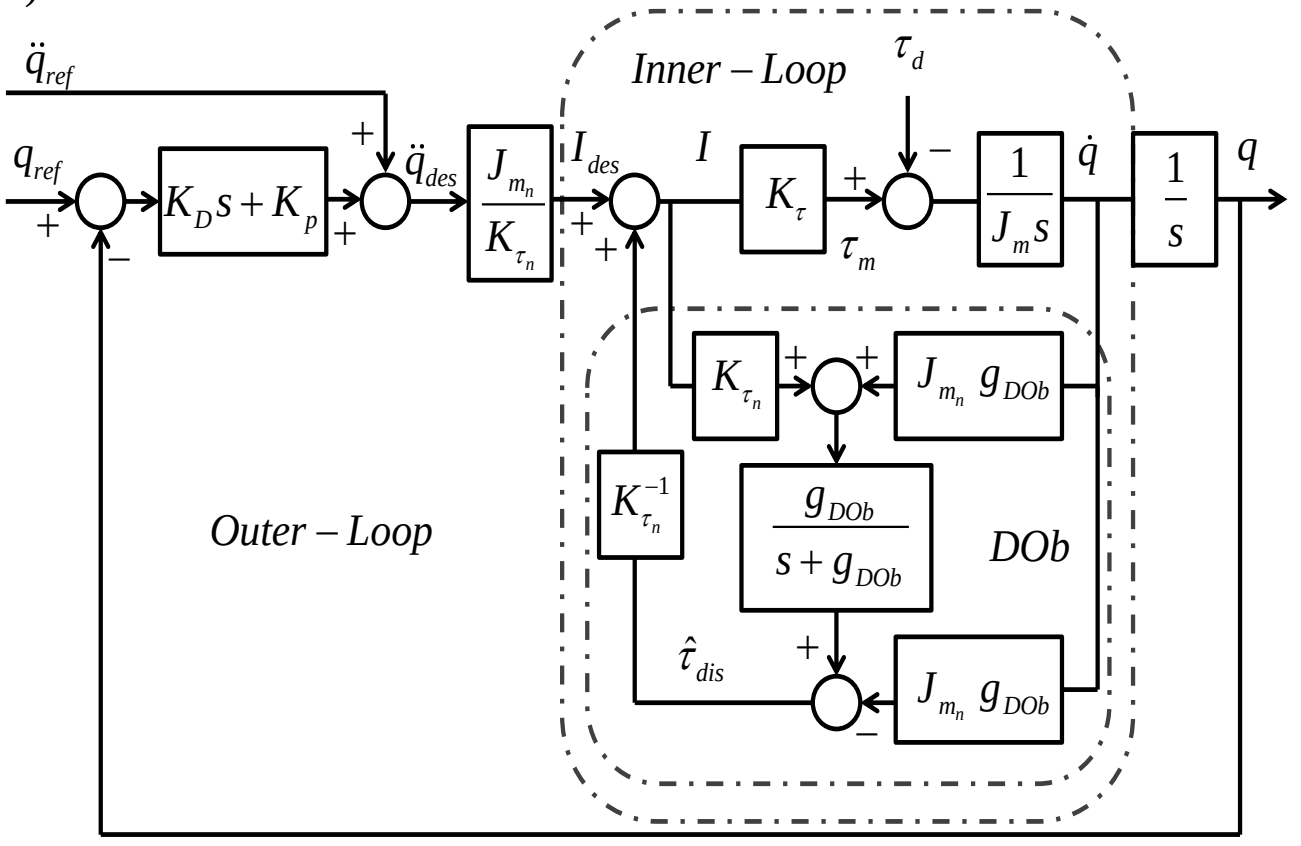

b) DOb-based robust position controller in the continuous-time domain.

Fig. 1: Block diagrams of the DOb and the robust motion controller in the continuous-time domain.

domains so that it is shown that continuous-time analysis methods fall-short in explaining the robust stability and performance of the DOb-based robust motion control systems implemented by computers and microcontrollers. The proposed discrete-time analysis shows that the robust stability and performance of the digital motion controller deteriorate as the bandwidth (i.e., robustness) and nominal inertia (i.e., phase margin) of the DOb increase. This explains why the digital robust motion controller becomes more sensitive to disturbances and unstable as the bandwidth and nominal inertia are increased in practice. To systematically synthesise a high performance digital robust motion controller, the upper bounds of the design parameters are analytically derived in this paper. Simulation results are given to verify the proposed analysis and synthesis methods.

The rest of the paper is organised as follows. The conventional DOb-based robust motion controller is presented in the continuous-time domain in section II. The stability and robustness of the DOb-based digital motion controller are analysed in the discrete-time domain in section III. New design constraints are derived to systematically synthesise the digital robust motion controller. Simulation results are given to verify the proposed analysis and synthesis methods in section IV. The paper ends with conclusion given in section V.

## II. DOb-based Robust Motion Controller in the Continuous-Time Domain

This section presents the DOb-based robust motion controller in the continuous-time domain. Block diagrams of the DOb and the robust motion controller are illustrated in the continuous-time domain in Fig. 1 [1, 17]. In this figure, the following apply:

| | |
|---|---|
| $J_m$ and $J_{m_n}$ | uncertain and nominal inertiae; |
| $K_\tau$ and $K_{\tau_n}$ | uncertain and nominal thrust coefficients; |
| $\tau_d$ and $\eta_V$ | disturbance and noise exogenous inputs; |
| $q$, $\dot{q}$ and $\ddot{q}$ | angle, velocity and acceleration; |
| $g_{DOb}$ and $g_v$ | bandwidths of the DOb and measurement; |
| $I$ | current of a DC motor; |
| $\tau_{dis}$ and $I_{dis}$ | disturbance torque and current; |
| $\hat{\bullet}$ | estimation of $\bullet$; |
| $\bullet_{des}$ and $\bullet_{ref}$ | desired $\bullet$ and reference $\bullet$; |
| $K_P$ and $K_D$ | proportional and derivative control gains; |
| $s$ | complex Laplace variable. |

The DOb-based robust motion controller has a 2-degrees-of-freedom control structure [1, 34]. While the robustness of the motion controller is improved via the DOb in the inner-loop, the outer-loop performance controller, i.e., PD controller, can independently adjust the performance of the position control system in the outer-loop. The inner- and outer- loop motion control structures are illustrated in Fig. 1b.

Let us first analyse the stability and robustness of the DOb by using Fig. 1a. The transfer functions between the exogenous inputs and acceleration are as follows:

When $g_v$ is infinite:

$$\ddot{q}(s)=\alpha\frac{s+g_{DOb}}{s+\alpha g_{DOb}}\ddot{q}_{des}(s)-\frac{1}{J_m}S_i(s)\tau_d(s)-sT_i(s)\eta_V(s) \quad (1)$$

where $S_i(s)=\dfrac{1}{1+L_i(s)}$ and $T_i(s)=\dfrac{L_i(s)}{1+L_i(s)}$ are the sensitivity and complementary sensitivity transfer functions in which $L_i(s)=\alpha\dfrac{g_{DOb}}{s}$ and $\alpha=\dfrac{J_{m_n}K_\tau}{J_m K_{\tau_n}}$ [17].

When $g_v$ is finite:

$$\ddot{q}(s)=\alpha\frac{(s+g_v)(s+g_{DOb})}{s^2+g_v s+\alpha g_v g_{DOb}}\ddot{q}_{des}(s)-\frac{1}{J_m}S_i(s)\tau_d(s)-sT_i(s)\eta_V(s) \quad (2)$$

where $\alpha$, $S_i(s)$ and $T_i(s)$ are same as defined in Eq. 1, however $L_i(s)=\alpha\dfrac{g_v g_{DOb}}{s(s+g_v)}$ [17].

Equations (1) and (2) show that a phase-lead/lag compensator is implicitly synthesised when the conventional DOb is used in the inner-loop of the robust motion controller. As $\alpha$ is increased, i.e., the nominal inertia is increased or the nominal thrust coefficient is decreased, the stability improves by increasing the phase margin of the robust motion control system. The inner-loop of the robust motion controller is stable for all values of the design parameters of $\alpha$ and $g_{DOb}$.

To analyse the robustness of the motion controller, let us apply Bode Integral Theorem to the DOb illustrated in Fig. 1a [18]. The Bode's integral equation is as follows:

When $g_v$ is infinite:

$$\int_0^\infty \log\left(\left|S_i(j\omega)\right|\right)d\omega=\pi\sum_k \mathrm{Re}\left(p^i_{u_k}\right)-\frac{\pi}{2}\lim_{s\to\infty}sL_i(s)=-\frac{\pi}{2}\alpha g_{DOb} \quad (3)$$

When $g_v$ is finite:

$$\int_0^\infty \log\left(\left|S_i\left(j\omega\right)\right|\right)d\omega = \pi\sum_k \mathrm{Re}\left(p_{u_k}^i\right) - \frac{\pi}{2}\lim_{s\to\infty} s\mathrm{L}_i\left(\mathrm{s}\right) = 0 \qquad (4)$$

where $\omega$ is frequency, $j^2 = -1$ is complex number and $\mathrm{Re}\left(p_{u_k}^i\right)$ is the real part of the $k^{th}$ right-half-plane pole of $\mathrm{L}^i\left(s\right)$ [18, 25].

Since the right hand side of Eq. (3) gets a lower value as $\alpha$ and/or $g_{DOb}$ are increased, the Bode's integral equation can hold without a high-sensitivity peak as the robustness and phase margin of the motion control system are improved. In other words, *waterbed effect* is not observed, and good robust stability and performance can be achieved for all values of the design parameters of $\alpha$ and $g_{DOb}$ when the DOb is synthesised by using ideal velocity measurement, i.e., $g_v$ is infinite. As the robustness against disturbances is improved by increasing either $\alpha$ or $g_{DOb}$, the peak of the sensitivity function increases to hold the Bode's integral equation given in Eq. (4). In other words, the robust motion controller may be subject to *waterbed effect* when a low-pass-filter is used in velocity measurement. This makes the robust motion control system more sensitive to disturbances at middle/high frequencies and degrades the robust stability and performance. However, the continuous-time analysis shows that the robust motion controller is stable for all values of the design parameters of $\alpha$ and $g_{DOb}$ [35].

The reader is invited to refer to [17, 35] for further details on the robustness analysis of the DOb in the continuous-time domain. Although continuous-time analysis methods provide good understanding for the asymptotic dynamic behaviours of the digital robust motion controller (e.g., the robustness against disturbances improves at low frequencies as the bandwidth of the DOb increases), they fall-short in explaining some dynamic responses. For example, it is a well-known fact that the digital robust motion controller exhibits oscillatory response and becomes unstable as the phase margin and robustness of the DOb are increased although Eqs. (1) – (4) show that the robust motion controller is stable for all values of the design parameters of $\alpha$ and $g_{DOb}$ [22–24]. Therefore, we may observe some unexpected dynamic responses when we analyse and synthesise the DOb-based digital robust motion controller in the continuous-time domain.

Let us now analyse the robust motion controller illustrated in Fig. 1b. The outer-loop's sensitivity and complementary sensitivity transfer functions are as follows:

$$S_o\left(s\right) = \frac{1}{1+L_o\left(s\right)} \quad \text{and} \quad T_o\left(s\right) = \frac{L_o\left(s\right)}{1+L_o\left(s\right)} \qquad (5)$$

where $L_o\left(s\right) = \alpha\dfrac{g_{DOb}s^2 + \left(s+g_{DOb}\right)\left(K_D s + K_P\right)}{s^3}$ when $g_v$ is infinite and $L_o\left(s\right) = \alpha\dfrac{g_v g_{DOB}s^2 + \left(s+g_v\right)\left(s+g_{DOB}\right)\left(K_D s + K_P\right)}{s^3\left(s+g_v\right)}$ when $g_v$ is finite.

Equation (5) shows that the dynamics of the DOb directly influences the stability and performance of the robust motion controller. For example, the following design constraint should hold to achieve stability [17].

$$\alpha^{-1} < 1 + g_{DOb}\frac{K_D}{K_p} + \frac{K_D}{g_{DOb}} + \frac{K_D^2}{K_p} \qquad (6)$$

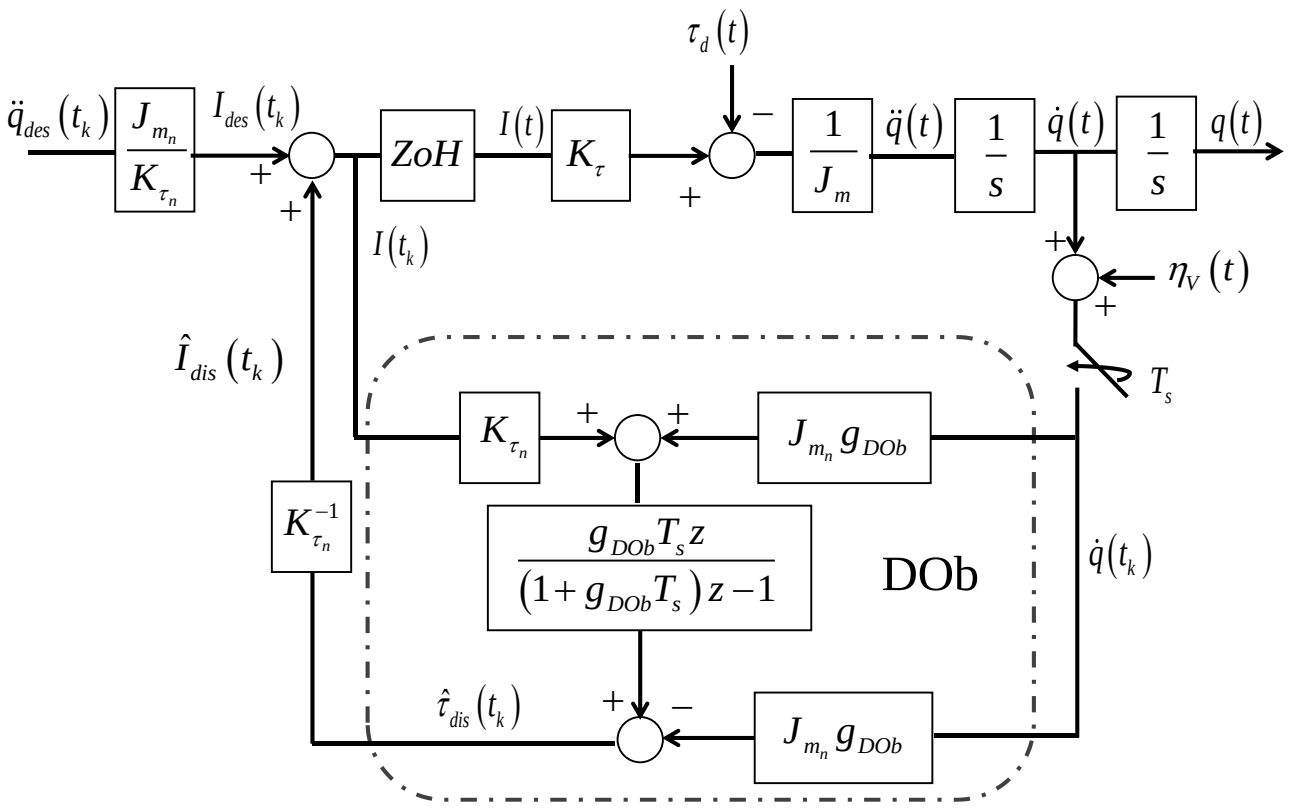

a) DOb in the discrete-time domain.

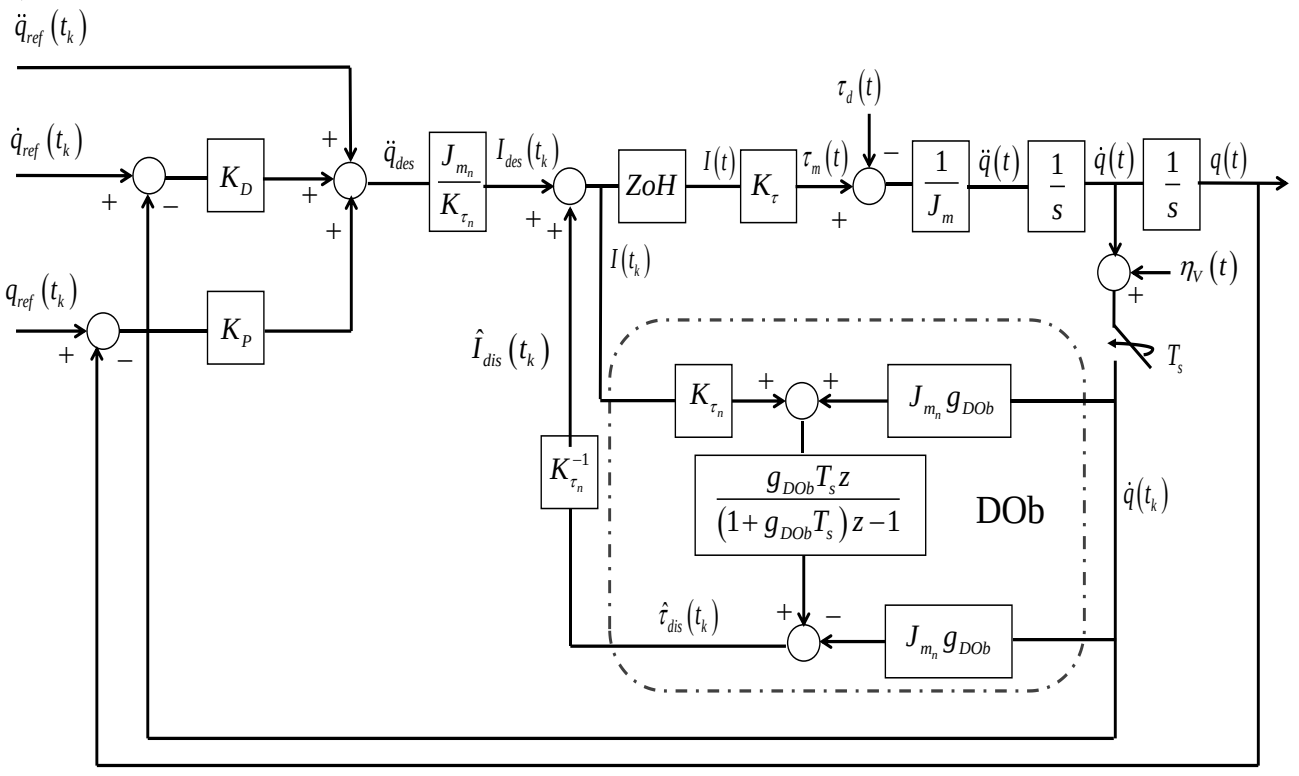

b) DOb-based robust position controller in the discrete-time domain.

Fig. 2: Block diagrams of the DOb and the robust motion controller in the discrete-time domain.

Equation (5) also shows that not only the DOb but also the outer-loop performance controller influences the robustness of the motion controller. As the control gains $K_p$ and $K_D$ are increased, the robustness against disturbances improves.

Although it is generally assumed that the robustness and performance of the DOb-based motion controllers can be independently adjusted in the inner- and outer- loop, respectively, Eq. (5) shows that this assumption is incorrect. The outer-loop controller can be used to tune the robustness, and the design parameters of the DOb can be used to tune the stability and performance of the motion controller [35]. In general, the bandwidth of the inner-loop is set higher than that of the outer-loop so that the influence of the disturbance estimation dynamics is suppressed [1, 17].

## III. DOb-based Robust Motion Controller in the Discrete-Time Domain

Block diagrams of the DOb and the robust motion controller are illustrated in the discrete-time domain in Fig. 2. In this figure, $T_s$ represents sampling time, $t$ and $t_k$ represent time in the continuous- and discrete- domains, respectively, $z$ represents a complex variable, and $ZoH$ represents zero-order-hold. The other parameters are defined earlier.

The inner-loop transfer functions between the exogenous inputs and the acceleration can be derived from Fig. 2a as follows:

$$\ddot{q} = \alpha\frac{\left(1+g_{DOb}T_s\right)z-1}{z-\left(1-\alpha g_{DOb}T_s\right)}\ddot{q}_{des} - \frac{1}{J_m}\mathrm{S_i}\left(z\right)\tau_d - \frac{\left(z-1\right)}{T_s}\mathrm{T_i}\left(z\right)\eta_V \qquad (7)$$

where $\mathrm{S_i}\left(z\right) = \dfrac{1}{1+\mathrm{L_i}\left(z\right)} = \dfrac{z-1}{z-\left(1-\alpha g_{DOb}T_s\right)}$ and $\mathrm{T_i}\left(z\right) = \dfrac{\mathrm{L_i}\left(z\right)}{1+\mathrm{L_i}\left(z\right)} = \dfrac{\alpha g_{DOb}T_s}{z-\left(1-\alpha g_{DOb}T_s\right)}$ are the inner-loop's sensitivity and complementary sensitivity transfer functions in which $L_i(z) = \frac{\alpha g_{DOb} T_s}{z-1}$.

Similar to the continuous-time analysis given in Eq. (1), Eq. (7) shows that a phase-lead (phase-lag) compensator is implicitly synthesised when the design parameters of the DOb are tuned by using $\alpha > 1/(1+g_{DOb}T_s)$ $(\alpha < 1/(1+g_{DOb}T_s))$. The phase-margin of the robust motion controller improves as $\alpha$ is increased. Since the sensitivity function gets smaller values at low frequency range, the robustness against disturbances can be improved by increasing either $\alpha$ or $g_{DOb}$. However, Eq. (7) shows that the inner-loop transfer functions become unstable when $\alpha g_{DOb}T_s > 2$, and the robust motion controller exhibits oscillatory response when $\alpha g_{DOb}T_s > 1$. In other words, neither $\alpha$ nor $g_{DOb}$ can be freely increased to improve the phase-margin and the robustness of the motion controller.

Let us employ Bode Integral Theorem to analyse the robustness of the DOb in the discrete-time domain [25]. The Bode integral equation of the DOb illustrated in Fig. 2a is as follows:

$$\int_{-\pi}^{\pi} \ln\left|S_i\left(e^{j\omega T_s}\right)\right| d\omega T_s = -2\pi \ln\left|1+\lim_{z\to\infty} L_i(z)\right| = 0 \quad (8)$$

To hold Eq. (8) within a limited frequency range (i.e., between $-\pi$ rad/s and $\pi$ rad/s), the peak of the sensitivity function increases as the robustness of the motion controller improves by increasing either $\alpha$ or $g_{DOb}$. In other words, the DOb-based digital robust motion controller is subject to *waterbed effect* even ideal velocity measurement ($g_v$ is infinite) is employed in the controller synthesis. As shown in section II, this dynamic behaviour of the digital robust motion controller cannot be deduced by conducting analysis in the continuous-time domain.

Let us consider the relation between the peaks of the sensitivity functions and the design parameters of the DOb in detail. The frequency responses of the inner-loop's sensitivity and complementary sensitivity transfer functions are as follows:

$$S_i(j\omega T_s) = \frac{\cos(\omega T_s) + j\sin(\omega T_s) - 1}{\cos(\omega T_s) + j\sin(\omega T_s) - (1-\alpha g_{DOb} T_s)} \quad (9)$$

$$T_i(j\omega T_s) = \frac{\alpha g_{DOb} T_s}{\cos(\omega T_s) + j\sin(\omega T_s) - (1-\alpha g_{DOb} T_s)} \quad (10)$$

When $2 > \alpha g_{DOb} T_s > 0$ (the stability constraint given in Eq. (7)), the maximum values of the sensitivity and complementary sensitivity functions appear at $\omega = T_s^{-1}(\pi + 2k\pi)$ rad/s as follows:

$$\left|S_i(j\omega T_s)\right|_{\max} = \left|S_i(j\pi)\right|_{\max} \frac{2}{\left|\alpha g_{DOb} T_s - 2\right|} \quad (11)$$

$$\left|T_i(j\omega T_s)\right|_{\max} = \left|T_i(j\pi)\right|_{\max} \frac{\alpha g_{DOb} T}{\left|\alpha g_{DOb} T_s - 2\right|} \quad (12)$$

If we assume that the sensitivity and complementary sensitivity functions satisfy $\left|S_i(j\omega T_s)\right|_{\max} \le 1/\Gamma_{S_i}$ and $\left|T_i(j\omega T_s)\right|_{\max} \le 1/\Gamma_{T_i}$, then the design constraints of the digital robust motion controller are derived as follows:

$$0 < \alpha g_{DOb} T_s \le 2\left(1-\Gamma_{S_i}\right) \quad (13)$$

$$0 < \alpha g_{DOb} T_s \le 2/\left(1+\Gamma_{T_i}\right) \quad (14)$$

where $0 < \Gamma_{S_i} < 1$ and $0 < \Gamma_{T_i} < 1$.

Let us now analyse the digital robust motion controller illustrated in Fig. 2b. For the sake of simplicity, the outer-loop performance controller is synthesised by using position measurement and Backward Euler method. The outer-loop transfer functions are derived from Fig. 2b as follows:

$$S_o(z) = \frac{1}{1+L_o(z)} \text{ and } T_o(s) = \frac{L_o(z)}{1+L_o(z)} \quad (15)$$

where $L_o(z) = C(z)C_i(z)G_p(z)$ in which $C(z) = K_P + K_D\frac{z-1}{T_s z}$, $C_i(z) = \alpha\frac{(1+g_{DOb}T_s)z-1}{z-(1-\alpha g_{DOb}T_s)}$ and $G_p(z) = \frac{T_s^2}{2}\frac{z+1}{(z-1)^2}$ [22].

Equation (15) similarly shows that the design parameters of the DOb may significantly influence the stability and performance of the digital robust motion controller. For example, the stability can be improved by tuning $\alpha > 1/(1+g_{DOb}T_s)$, i.e., designing $C_i(z)$ as a phase-lead compensator. Besides, the robustness against disturbances can be improved by properly tuning the outer-loop performance controller so that lower values of the sensitivity function is obtained at low-frequencies.

## IV. Simulation Results

In this section, simulation results are given to verify the proposed analysis and synthesis methods.

Let us start with the continuous-time analysis methods. Figure 3 illustrates the frequency responses of the sensitivity and complementary sensitivity transfer functions when the nominal inertia and the bandwidth of the DOb are set at different values. It is clear from this figure that good robust stability and performance can be achieved for all values of the design parameters of $\alpha$ and $g_{DOb}$ when ideal velocity measurement is employed in the DOb synthesis. However, the robust motion controller is subject to *waterbed effect* and the peak of the sensitivity and complementary sensitivity function increases as the robustness against disturbances is improved when $g_v$ is finite.

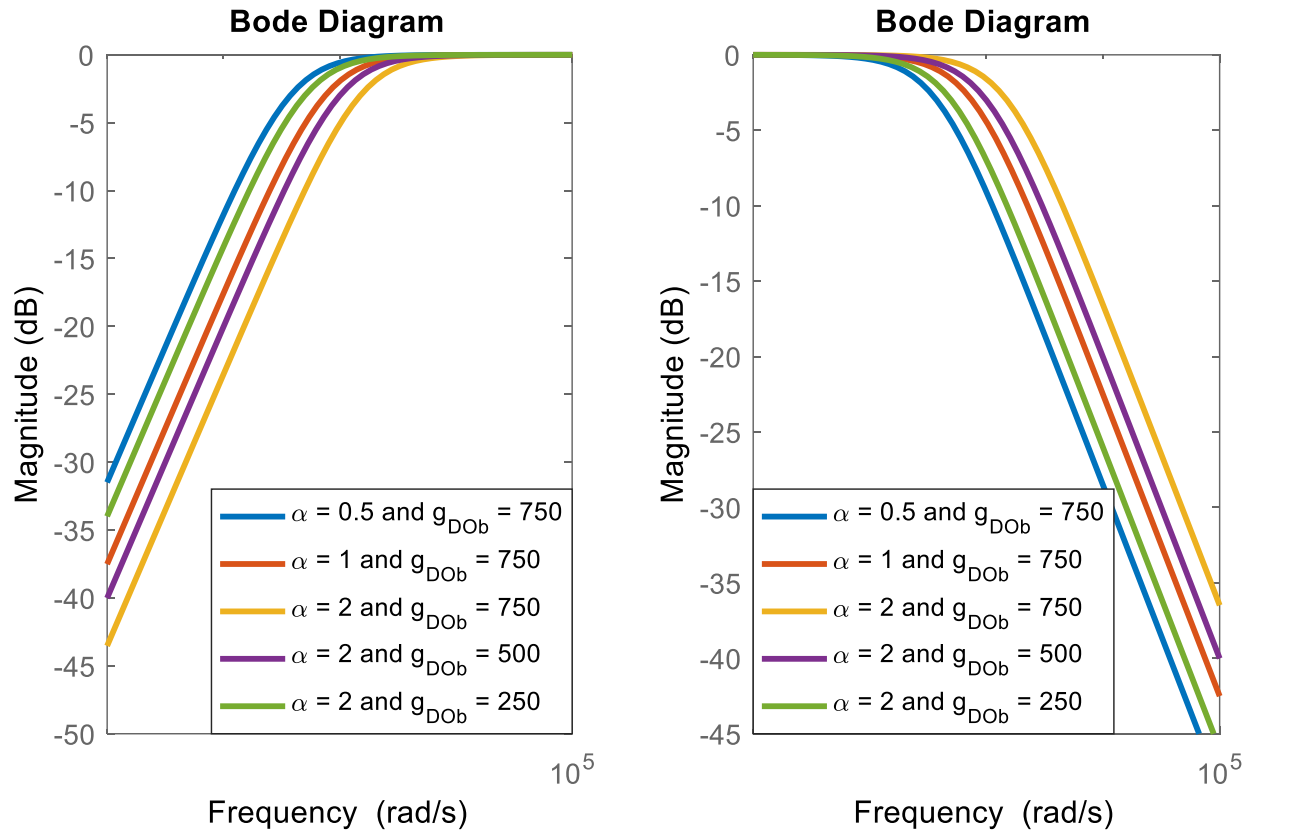

a) When $g_v$ is infinite.

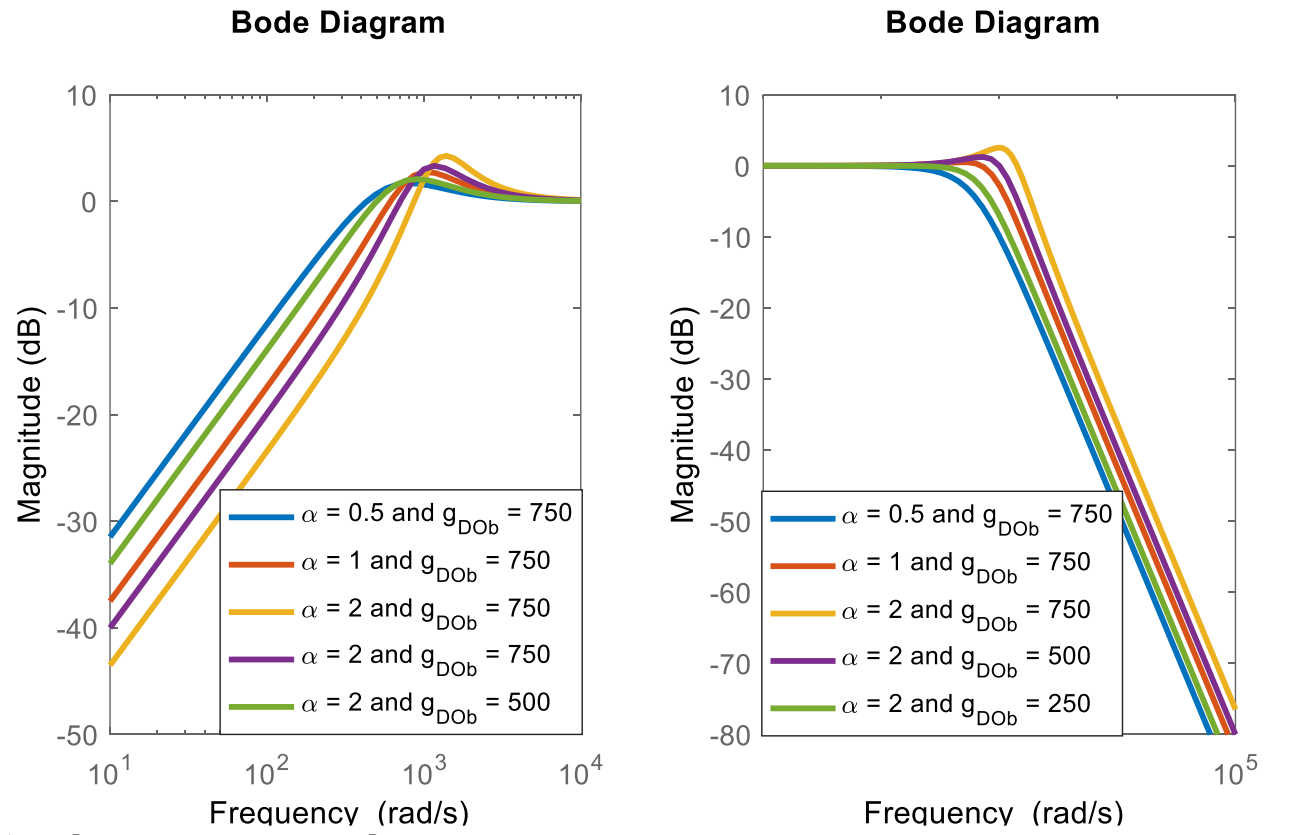

b) When $g_v$ is 1000 rad/s.

Fig.3. Sensitivity (left-figures) and complementary sensitivity (right-figures) functions' frequency responses in the continuous-time domain.

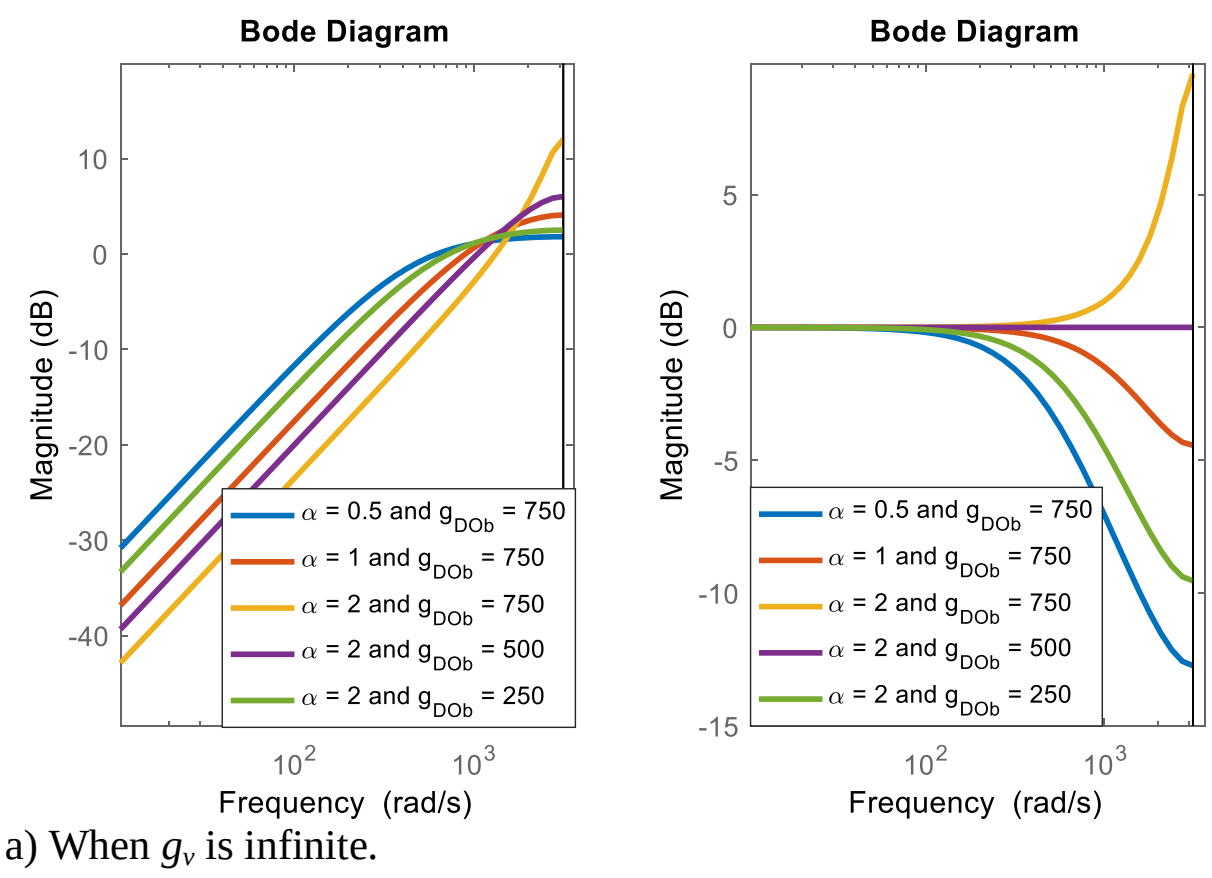

a) When $g_v$ is infinite.

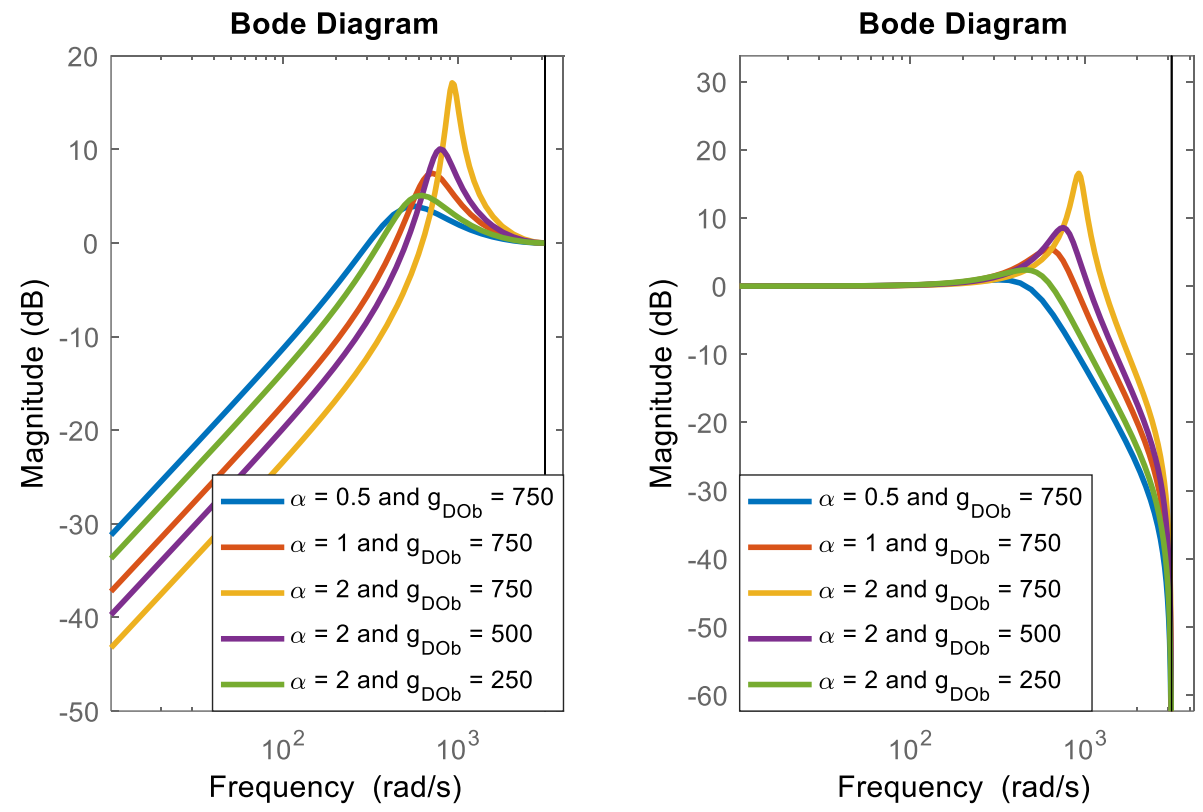

b) When $g_v$ is 1000 rad/s.

Fig.4.Sensitivity (left-figures) and complementary sensitivity (right-figures) functions' frequency responses in the discrete-time domain when $T_s$ = 1ms.

Let us now analyse the DOb-based robust motion controller in the discrete-time domain. Figure 4 illustrates the frequency responses of the sensitivity and complementary sensitivity transfer functions for different values of $\alpha$ and $g_{DOb}$. As the robustness against disturbances is improved by increasing the bandwidth of the DOb or the phase margin is improved by increasing/decreasing the nominal inertia/thrust coefficient, the peaks of the sensitivity and complementary sensitivity transfer functions increase. This makes the robust motion controller more sensitive to disturbances at high frequencies such as noise and degrades the robust stability and performance.

Let us now consider the stability of the robust motion controller. Figure 5 illustrates the root-loci of the motion control system with respect to $\alpha$ and $g_{DOb}$ in the continuous- and discrete- time domains. This figure shows that the robust motion controller becomes unstable for small values of $\alpha$ and the stability is improved with phase-lead effect when $\alpha$ is increased. However, only the discrete-time analysis shows that the robust motion controller becomes unstable for high-values of $\alpha$. Similarly, Fig. 5c shows that the digital robust motion controller becomes unstable as the bandwidth of the DOb is increased.

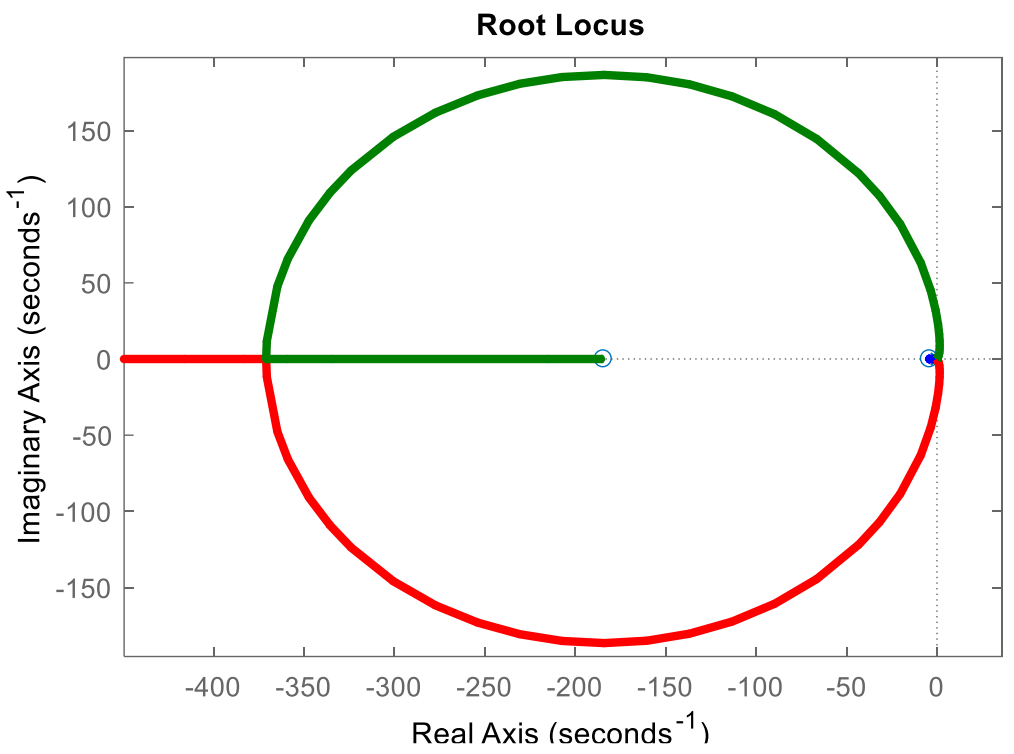

a) Root-locus with respect to $\alpha$ in the continuous-time domain.

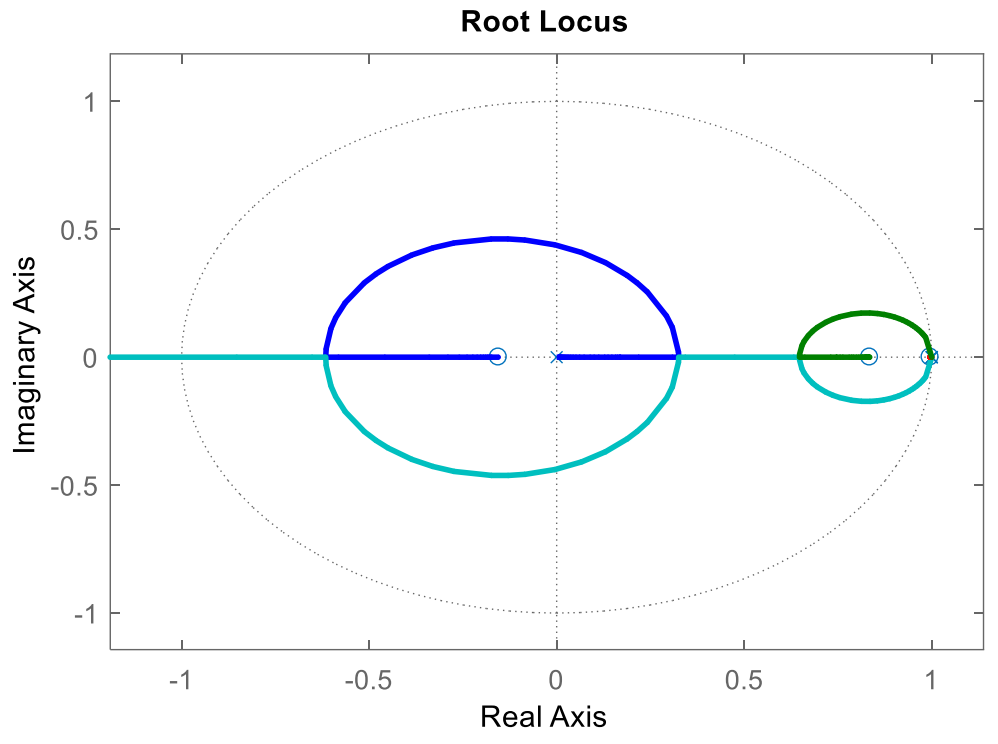

b) Root-locus with respect to $\alpha$ in the discrete-time domain.

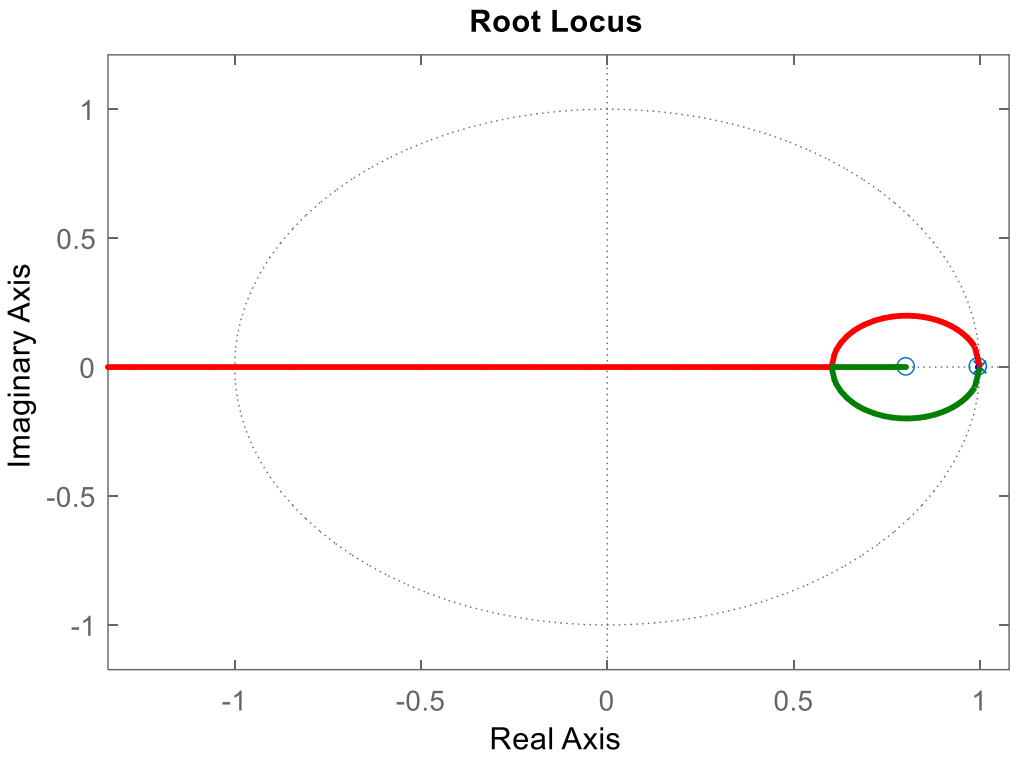

c) Root-locus with respect to $g_{DOb}$ in the discrete-time domain. $\alpha = 0.01$.

Fig. 5. Root-loci of the robust motion controller. The parameters of the simulations are $J_m$ = 0.003, $K_\tau$ = 0.25, $g_{DOb}$ = 750, $K_P$ = 1000, $K_D$ = 250, and $T_s$ = 1ms.

Last, let us present robust position control results. In this simulation, it is assumed that a servo system is affected by internal and external disturbances. Figure 6 shows that the robust position controller can precisely follow regulation and trajectory tracking control references by supressing internal and external disturbances when it is synthesised by employing the design constraints derived in section III. When the design constraints are not satisfied (e.g., blue curves for $\alpha \ll 1$ and grey curves for $\alpha g_{DOb} T_s > 1$), oscillatory and unstable responses are observed.

## V. CONCLUSION

In this paper, the stability and robustness of the DOb-based motion control systems are analysed in the continuous- and discrete- time domains. Although continuous-time analysis methods are useful to explain the asymptotic dynamic behaviours of the DOb implemented by computers and microcontrollers, they fall-short in explaining the robust stability and performance of the digital motion controller. For example, continuous-time analysis methods cannot explain why the digital robust motion controller becomes unstable as the phase-margin and robustness are improved by increasing $\alpha$ and $g_{DOb}$. New design constraints on the nominal plant model and the bandwidth of the DOb are analytically derived in discrete-time. One can systematically synthesise a high-performance digital robust motion controller by employing the proposed design constraints. To achieve good robust stability and performance, this paper recommends discrete-time analysis and synthesis methods for the DOb-based robust motion controllers implemented by computers.

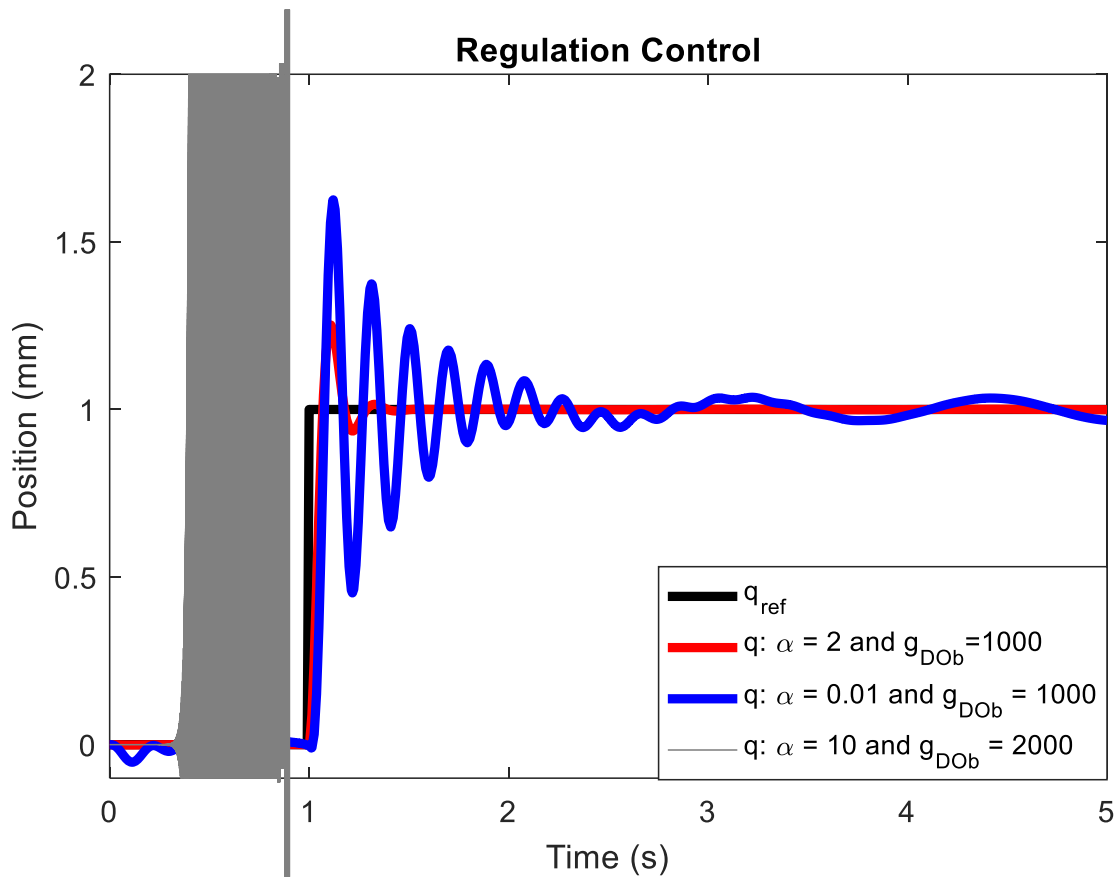

a) Regulation control.

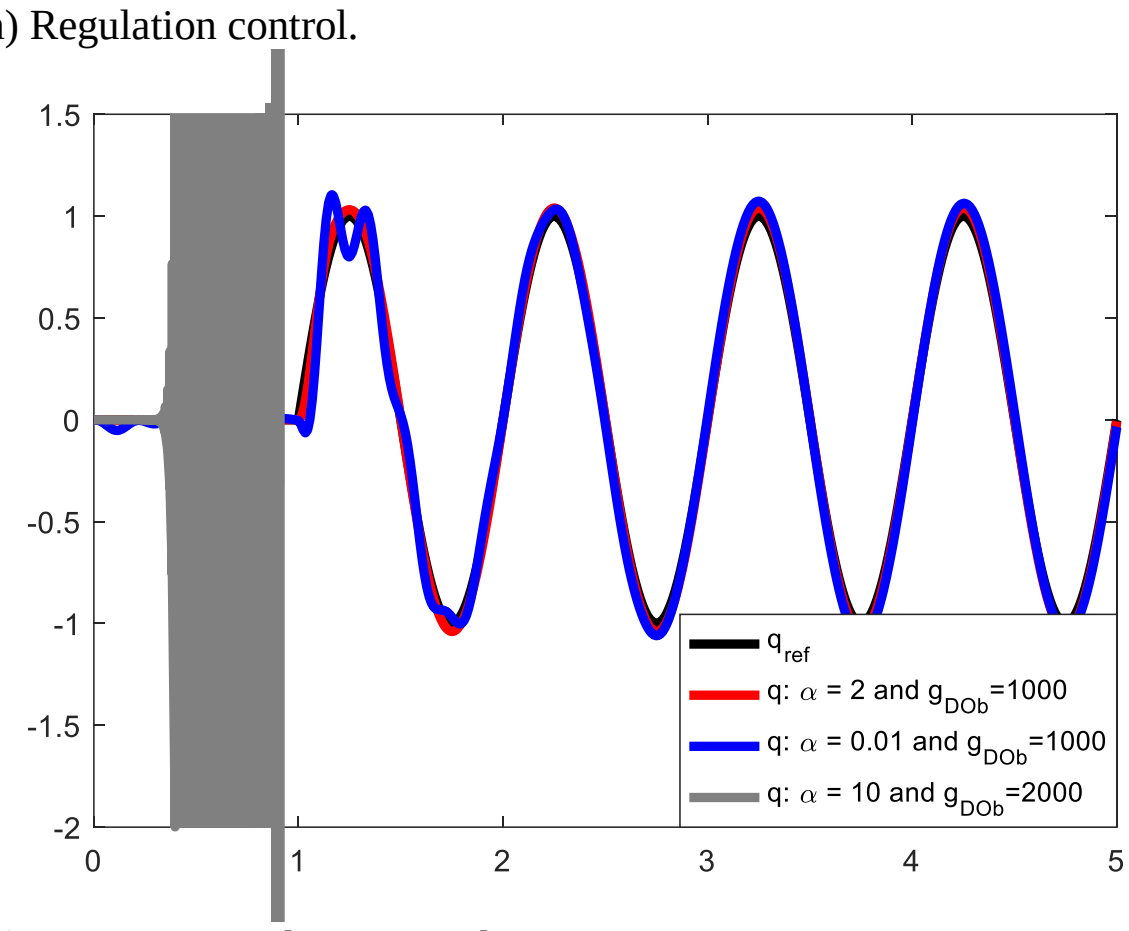

b) Trajectory tracking control.

Fig. 6. Position control responses. The parameters of the simulations are $J_m$ = 0.003, $K_\tau$ = 0.25, $K_P$ = 1000, $K_D$ = 25, and $T_s$ = 0.1ms.